\pdfoutput=1
\documentclass[aps,prc,floatfix,reprint,longbibliography,nofootinbib,showkeys]{revtex4-1}

\usepackage{amsmath,amssymb}    
\usepackage{graphicx}   
\usepackage{color}
\usepackage{siunitx}  
\usepackage[colorlinks,linkcolor=blue,citecolor=blue]{hyperref}   
\usepackage{bm} 
\usepackage{import}

\def\L{{\mathcal{L}}}
\def\P{{\mathcal{P}}}

\def\x{{\bm x}}

\def\p{{\bf p}}
\def\rmd{{\rm d}}

\newcommand{\rme}{{\rm e}}

\newcommand\nda{\end{align}}
\def\Eq#1{Eq.~(\ref{#1})}
\def\Eqs#1{Eqs.~(\ref{#1})}

\def\Fig#1{Fig.~\ref{#1}}

\def\Ref#1{Ref.~\cite{#1}}

\def\t{\tilde}

\advance\parskip 1.9pt
\advance\voffset -0.2in

\def\be{\begin{equation}}
\def\ee{\end{equation}}
\def\bg{\begin{eqnarray}}
\def\nd{\end{eqnarray}}

\def\Refs{\cite}

\begin{document}


\title{Analytical attractor for Bjorken flows}


\author{Jean-Paul Blaizot}
\email[]{jean-paul.blaizot@ipht.fr}
\affiliation{
	Institut de Physique Th{\'e}orique, Universit\'e Paris Saclay, 
        CEA, CNRS, 
	F-91191 Gif-sur-Yvette, France} 
\author{Li Yan}
\email[]{cliyan@fudan.edu.cn}
\affiliation{Key Laboratory of Nuclear Physics and Ion-Beam Application (MOE) \& Institute of Modern Physics\\
Fudan University, 220 Handan Road, 200433, Yangpu District, Shanghai, China}

\date{\today}

\begin{abstract}
We present an analytic solution of a simple set of equations that govern the expansion of boost-invariant plasmas of massless particles.  These equations describe,  approximately, the early time, collisionless regime, and the transition to hydrodynamics at late time. Their mathematical structure encompasses all versions of second order hydrodynamics when applied to Bjorken flows.  The analytic solution provides an explicit expression for the attractor solution that connects the collisionless regime to hydrodynamics. It also constitutes a neat example of an application of the theory of resurgence in asymptotic series. 

\end{abstract}
\maketitle

The success of the fluid dynamical description of ultra-relativistic heavy ion collisions has triggered recently a lot of theoretical work \cite{romatschke2017relativistic,Florkowski:2017olj}.   While hydrodynamics is often viewed as an effective theory for long wavelength modes with microscopic degrees of freedom near local equilibrium, implying small gradients and small mean free paths, a number of recent studies suggest that it may work even when these conditions are not fully satisfied.  It was found for instance that viscous corrections can account for the large pressure anisotropy in the longitudinally expanding plasma  formed in heavy ion collisions,  for either a strongly coupled system in holography \cite{Heller:2011ju}, or a weakly coupled system in kinetic theory \cite{Kurkela:2015qoa}. Thus the general question of  how hydrodynamical behavior  emerges in various systems has become a prominent one. In this context, an important progress has been  the  identification of attractor solutions in dynamical equations whose late time behavior is hydrodynamical \cite{Heller:2015dha,Romatschke:2017vte,Denicol:2017lxn,Kurkela:2019set}.  Besides its physical relevance, this  issue also connects with beautiful mathematical developments in the theory of resurgence in asymptotic series \cite{Aniceto:2018bis,Basar:2015ava}.

This letter aims to contribute to this general discussion by providing an explicit analytical solution for a simple set of equations that describes the transition from kinetics to hydrodynamics for a rapidly expanding plasma of massless particles.  The system that we consider is an idealization of  the early stage of a high-energy heavy-ion collision, where matter expands longitudinally along the collision axis in a boost invariant fashion, the so-called Bjorken expansion \cite{Bjorken:1982qr}.  Our starting point is a simple kinetic equation, $p^\mu\partial_\mu f(t,\x,\p) ={\cal C}[f]$, where $f$ denotes a distribution function for massless particles, and ${\cal C}[f]$ is a collision term treated in the relaxation time approximation \cite{Baym:1984np}, a widely used  approximation \cite{Florkowski:2017olj}. 

It has been shown in previous works \cite{Blaizot:2017ucy} that the solution of  the kinetic equation  can be accurately reproduced by a set of coupled equations for the two independent components of the energy-momentum tensor, the  energy density $\varepsilon=\L_0$ and the viscous pressure, expressed as the difference between the longitudinal and transverse pressures $\P_L-\P_T=\L_1$.  Owing to the symmetries of the Bjorken expansion, $\L_0 $ and $\L_1$ depends only on the proper time $\tau$, and obey the coupled equations  
\begin{subequations}
\label{eq:Lequ}
\begin{align}
\label{eq:Lequa}
\frac{\partial \L_0}{\partial \tau} =& - \frac{1}{\tau}(a_0 \L_0 + c_0 \L_1)\,, \\
\label{eq:Lequb}
\frac{\partial \L_1}{\partial \tau} =& - \frac{1}{\tau}(a_1 \L_1 + b_1 \L_0) - \frac{\L_1}{\tau_R}\,,
\end{align}
\end{subequations}
where $\tau$ is the proper time. 
The  coefficients, $a_0=4/3$, $a_1=38/21$, $b_1=8/15$, and $c_0=2/3$,  are pure numbers whose values are fixed by the geometry of the expansion. The last term, proportional to the collision rate $1/\tau_R$,  isolates in a transparent way the effect of the collisions. We shall allow $\tau_R$ to depend on  $\tau$ (see below).  Eqs.~(\ref{eq:Lequ}) are the first in an infinite hierarchy of equations for moments defined by $
\L_n\equiv \int d^3 \p \,p P_{2n}(p_z/p) f_p(t,\x, \p),
$
with $P_n(x)$ a Legendre polynomial and $p=|\p|$ \cite{Blaizot:2017ucy} \footnote{These moments $\L_n$, introduced in \cite{Blaizot:2017lht}, are distinct from those most commonly used  (see e.g. \cite{Strickland:2018ayk}). A generalization is presented in \cite{Behtash:2019txb}.}. The goal of this paper is to present an exact analytical solution of Eqs.~(\ref{eq:Lequ}).

 Without the last term in Eq.~(\ref{eq:Lequb}), Eqs.~(\ref{eq:Lequ})  describe (approximately) free streaming: the moments evolve as power laws governed by the eigenvalues of the linear  system. The collision term in Eq.~(\ref{eq:Lequb}) produces a damping of  $\L_1$ and drives the system towards isotropy, a prerequisite for local equilibrium. When $\L_1=0$, the system behaves as in ideal hydrodynamics $\L_0\sim \tau^{-a_0}$. 

However there is more in the approach to hydrodynamics than the simple vanishing of $\L_1$. Indeed, the term $-\L_1/\tau_R$ in Eq.~(\ref{eq:Lequb}) suggests the possible presence of exponential terms in the solution, in addition to the power laws originating from the expansion. Such exponential contributions would spoil the hydrodynamic gradient expansion, which, in the present context, is an expansion in inverse powers of $\tau$. However,  these contributions cancel  thanks to the  relation $\L_1/\tau_R \simeq -b_1 \L_0/\tau $  valid at late time \cite{Blaizot:2017ucy}. This relation, in fact, happens to be the leading order constitutive equation relating the viscous pressure to the viscosity, with $\eta=(b_1/2)\varepsilon\tau_R$ being the standard value of the viscosity. We shall return to these exponential contributions later in this letter. 

The connection between Eqs.~(\ref{eq:Lequ}) and hydrodynamics is tight.  Eq.~(\ref{eq:Lequa})  translates the conservation of the energy momentum tensor, $\partial_\mu T^{\mu\nu}=0$, for Bjorken flow.  An equation similar to Eq.~(\ref{eq:Lequb}) was introduced by Israel and Stewart \cite{Israel:1979wp} to overcome  limitations of the Navier Stokes equation in the relativistic context. In fact, as discussed in  \cite{Blaizot:2019scw,Blaizot_Yan}, if one adjusts properly the coefficients $a_1$, $b_1$ and the relaxation time $\tau_R$ ($a_0$ and $c_0$ are fixed by energy-momentum conservation, as just mentioned), all the known formulations of second order hydrodynamics (e.g.~\cite{Baier:2007ix,Denicol:2012cn}) can be mapped, in the context of Bjorken flows, into Eqs.~(\ref{eq:Lequ}), and  share therefore the same mathematical structure. Indeed, the authors of Ref.~\cite{Denicol:2017lxn} (see also \cite{Jaiswal:2019cju}),  using some version of  Israel-Stewart  hydrodynamics, obtained a  solution that is a particular case of that presented here.

The use of a time dependent relaxation time $\tau_R(\tau)$ allows us to capture the qualitative features of more realistic calculations. The solution presented in this letter holds for a  time dependence of the form  $\tau_R \sim \tau^{1-\Delta}$,
with $\Delta$ constant\footnote{Fixing the units requires an additional time scale $\tau_1$, which can be chosen as the time at which the expansion rate equals the collision rate, that is $\tau_1=\tau_R(\tau_1)$.} \cite{Heller:2018qvh}.  Commonly used are a constant $\tau_R$  ($\Delta=1$), or $\tau_R\sim T^{-1}$ with $T$ the effective temperature ($\Delta\approx 2/3$). Note that as long as $\Delta>0$, the expansion rate decreases faster than the collision rate, and the system is driven to hydrodynamics at late time. When $\Delta<0$, the expansion eventually overcomes the collisions and the system evolves towards the free streaming regime. The limiting case $\Delta=0$ 
mimics the system of hard sphere scatterings recently studied in  \cite{Denicol:2019lio}, and is also close to more realistic QCD kinetics with  2-to-2 scatterings \cite{Blaizot:2019dut}. In this case,  collisional effects perfectly balance those of the expansion, resulting in a stationary state that differs from hydrodynamics. More sophisticated time dependence have also been considered \cite{Chattopadhyay:2019jqj}. A detailed description of these physical situations and the corresponding solutions as a function of $\Delta$ will be presented in a forthcoming publication \cite{Blaizot_Yan}.  In this letter we focus on the solution for  $\Delta>0$.

To proceed, it  is convenient to measure the time in units of  the instantaneous relaxation time and define the dimensionless variable
$
\label{eq:wdef}
w\equiv {\tau}/{\tau_R(\tau)}\sim \tau^{\Delta}, 
$
\begin{figure}
\begin{center}
\includegraphics[width=0.42\textwidth] {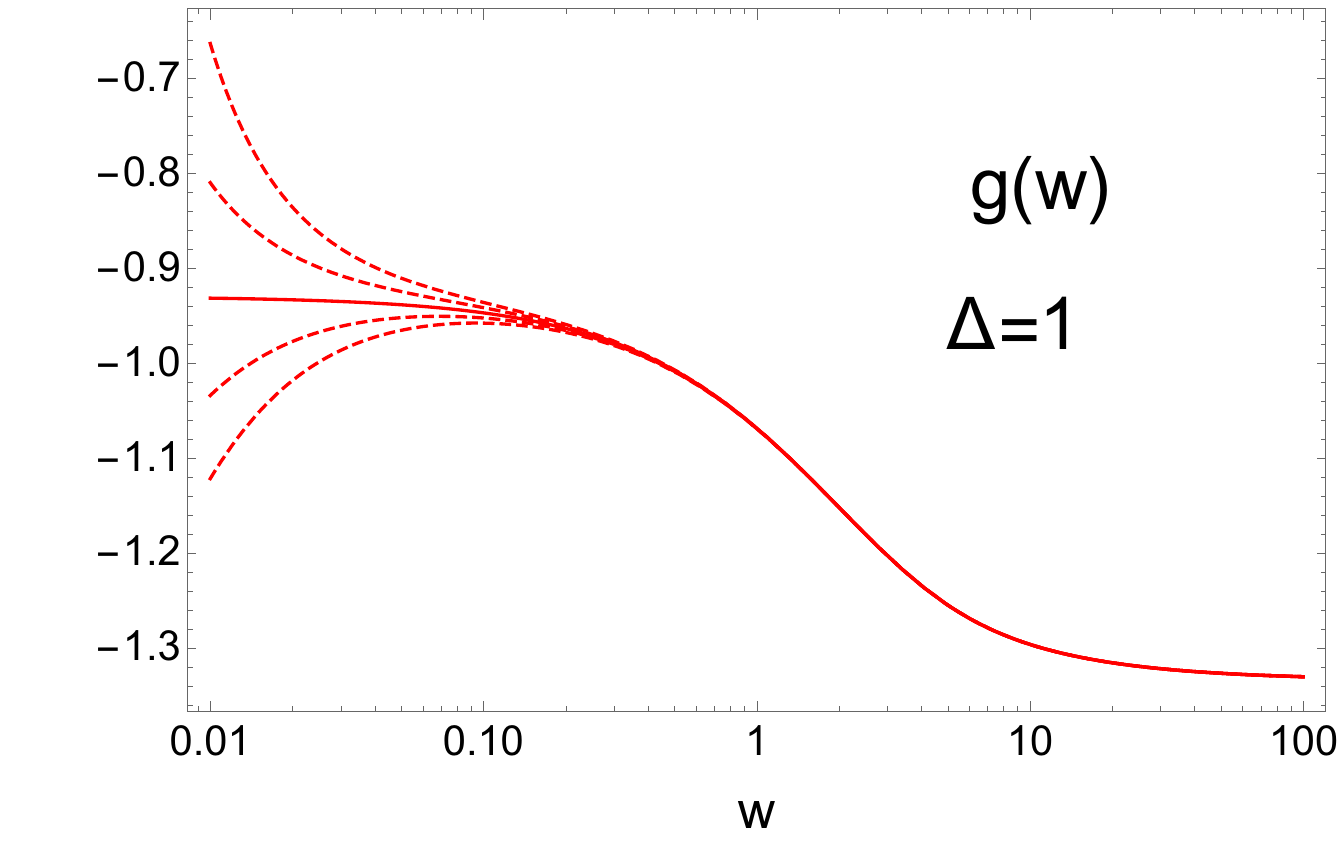}
\caption{ (Color online) 
Plot of $g(w)$ obtained from \Eq{eq:gsol2} for various initial conditions set up at the initial value $w_0=0.01$.  The solid line represents the attractor  joining the free streaming fixed point $g_+$ at small $w$ to the hydrodynamic fixed point $g=-4/3$ at large $w$. 
\label{fig:solution}
}
\end{center}
\end{figure}
which plays here the role of an inverse Knudsen number. 
We also define
\be
g(w) \equiv \frac{\tau}{\L_0}\frac{\partial \L_0}{\partial \tau}=-1-\frac{\P_L}{\epsilon}.
\ee
This quantity $g(w)$ may be viewed as the exponent of  the power laws obeyed by the energy density at early or late times. It is also a measure of the pressure asymmetry. In particular, the second relation, which follows easily from Eq.~(\ref{eq:Lequ}), shows that in the free streaming regime where $\P_L=0$, $g=-1$, while in the hydrodynamical regime where $\P_L=\varepsilon/3$, $g=-4/3$.  In terms of $g(w)$, \Eqs{eq:Lequ} become a first order nonlinear ODE,
\be
\label{eq:dif_gw}
\Delta\frac{\rmd g}{\rmd \ln w}+g^2+\left(a_0+a_1+w\right)g+a_1a_0-c_0b_1+ a_0 w=0\,.
\ee
In the absence of collisions, this non linear equation has two fixed points, that we refer to as unstable ($g_-$) and stable ($g_+$) free streaming fixed points, whose values coincide with the eigenvalues of the linear system (\ref{eq:Lequ}). Numerically, $g_+=-0.929$, $g_-=-2.213$ \footnote{The fact that $g_+$ is not exactly $-1$ is an effect of the two moment truncation \cite{Blaizot:2019scw}.}.  This fixed point structure continues to play a role when collisions are switched on \cite{Blaizot:2019scw}: The unstable fixed point moves to large negative values, while the stable fixed point $g_+$ evolves adiabatically\footnote{When $\Delta\to 0$ in Eq.~(\ref{eq:dif_gw}) an approximation akin to the adiabatic approximation becomes accurate, and remains so for $\Delta\lesssim 1$.} \cite{Brewer:2019oha} to the hydrodynamic fixed point, $g_*= -4/3$. The location of this ``pseudo fixed point'' as $w$ runs from $0$ to $\infty$ corresponds (approximately) to what has been  dubbed ``attractor''  \cite{Heller:2015dha}.  Here the attractor\footnote{For a definition of the notion of attractor more in line with that in use in the literature on dynamical systems we refer the reader to \cite{Behtash:2019txb,behtash2020transasymptotics} and references therein.} is understood  as the particular solution of Eq.~(\ref{eq:dif_gw}), $g_{\rm att}(w)$,  that connects  $g_+$ as $w\to 0$ to $g_*$ as $w\to\infty$.  As was shown  in \cite{Blaizot:2019scw} the fixed point structure at small $w$ is only moderately affected by the higher moments ($\L_n$, with $n\ge 2$),  which explains why the two-moment truncation (\ref{eq:Lequ}) is a good approximation to the solution of the full kinetic equation for the lowest two moments\footnote{The main effect of the higher moment is to bring $g_+$ to its correct value, thereby eliminating the regions of the attractor corresponding to negative longitudinal pressure.}. Finally, in view of the close relation between the coupled equations (\ref{eq:Lequ}) and hydrodynamics, the same fixed point structure plays a prominent role in second order hydrodynamics.

 By setting $g(w)+a_0=\Delta(b-a)-w+wy'(\bar w)/y(\bar w)$,
 with $b$ and $a$ constants, and $\bar w=w/\Delta$, one transforms Eq.~(\ref{eq:dif_gw}) into a linear, second order ODE:
\be\label{Kummer}
\bar w^2y''+y'(b-\bar w)-ay=0,\qquad  (y'=\rmd y/\rmd \bar w).
\ee
The solutions $y(\bar w)$ of this equation (Kummer's equation) are confluent hypergeometric functions $M(a,b,w/\Delta)$ and $U(a,b,w/\Delta)$ \cite{10.5555/1098650}. The parameters $a$ and $b$ can take one of two possible sets of values, 
$
a_\pm = 1 - \frac{g_\mp + a_0 }{\Delta}$, and $
b_{\pm} = 1\pm\frac{g_+-g_-}{\Delta}$. However, it can be shown that the physically meaningful solution  corresponds to the choice $a_+, b_+$\footnote{One can show for instance that only this choice leads to an increase of entropy in the late time hydrodynamic regime \cite{Blaizot_Yan}.}. Thus, from now on we shall simply set $a_+=a,b_+=b$ and write the general solution for $g(w)$ in the form
\begin{align}
\label{eq:gsol2}
g(w)& = g_+ 
-w \nonumber\\
+& a w
\frac{\frac{1}{b}{M\left(1+a,1+b,\frac{w}{\Delta}\right)}- A U\left(1+a,1+b,\frac{w}{\Delta}\right)}
{M\left(a,b,\frac{w}{\Delta}\right)+ A U\left(a,b,\frac{w}{\Delta}\right)}\,,
\end{align}
where $A$ is a constant  to be fixed by the initial condition\footnote{Note that a single initial condition is required for the three versions of the equations of motion, Eqs.~(\ref{eq:Lequ},\ref{eq:dif_gw},\ref{Kummer}).}. 
 \Eq{eq:gsol2} 
is the exact solution to Eq.~(\ref{eq:dif_gw}) for all $\Delta>0$. Since the structure of the solution does not change much as $\Delta$ varies, we set from now on $\Delta=1$.\footnote{Solutions for  values of $\Delta>0$ other than 1 can in fact be deduced from the solution for $\Delta=1$ via a suitable rescaling of the parameters of the equation. See~\cite{Blaizot_Yan} for details.} An illustration of the solution for various initial conditions is provided in \Fig{fig:solution} for  five values of $A$. 
As can be seen, the various solutions eventually 
merge onto the solid line, the attractor solution which  connects  $g_+$ to $g_*$. This attractor solution is obtained for $A=0$, viz.
\be
\label{eq:expgatt}
g_{\rm att}(w)=g_+ - w+w
\frac{a M(1+a,1+b,w)}{b M(a,b,w)}\,.
\ee

 This attractor solution possesses a well behaved expansion at small $w$. Physically, such an expansion  represents perturbative corrections to free-streaming  due to collisions,  and it emerges from Eqs.~(\ref{eq:Lequ}) as one  expands in powers of  the last term in Eq.~(\ref{eq:Lequb}).  When the parameters $a$ and $b$ differ from  non-negative integers, which is the case here, the function $M(a,b,w)$ is analytic and expressible as a convergent series in $w$.  However, in contrast to the function $M$ itself, the series  $
g^{\rm att}(w) = \sum_{n=0} \gamma_n^{\rm att} w^n$, with $\gamma_0^{\rm att}=g_+$,  has a finite radius of convergence determined by the location of the zero of $M(a,b,w)$ in the denominator of Eq.~(\ref{eq:expgatt}) that is closest to the origin. The Borel sum of the series, after analytic continuation, converges to the exact function, as it has been verified numerically in \cite{Kurkela:2019set}.

When $A\ne0$, the solution (\ref{eq:gsol2}) involves the function $U(a,b,w)$ which contains a singular contribution $\sim w^{-b}$ at the origin \cite{10.5555/1098650}.  The small $w$ expansion becomes then more intricate and takes the form of a trans-series  \cite{Behtash:2019txb} 
\be
g(w) = \sum_{m=0} w^{m(b-1)} \sum_{n=0} \gamma_{n}^{(m)} w^n\,,
\ee
where  the expansion coefficient $\gamma_{n}^{(m)}$ for $m>0$ depends on $A$. We note that the ratio $\gamma^{(0)}_n/\gamma^{(0)}_{n+1}$ approaches the value  $w\approx 0.3314$ corresponding to the finite radius of convergence of the $m=0$ series. Furthermore, $\gamma_{0}^{(0)}=g_-$ irrespective of the value of $A$. This implies that  all solutions (but the attractor) start from the unstable free-streaming fixed point at $w\to 0^+$, even though at any small but finite $w$ one can find a value of $A$ such that  a given solution can take any desired value.

At large $w$, a solution $g(w)$ of  \Eq{eq:dif_gw} can be obtained  as an expansion in powers of $1/w$, giving rise to the hydrodynamic gradient expansion,
\be
\label{eq:hydroexp}
g_{\rm hydro}(w) = \sum_n f_n w^{-n}. 
\ee
In early works, the coefficients $f_n$ were determined iteratively, their observed factorial growth reflecting the asymptotic character of the expansion~\cite{Heller_2013}. The knowledge of these coefficients could then be exploited in a Borel summation. The Borel sum is the inverse Laplace transform, 
\be\label{Bsum}
\tilde g(w)=\int_{\cal C} \rmd t\, \rme^{-t} {\cal B} [g(t/w)],
\ee 
where ${\cal B}[g(t)]$ is the Borel transform of the original series, viz.
\be
\label{eq:Boreltransf}
{\cal B}[g(t)]=\sum_n \frac{f_n t^n}{n!},
\ee 
and ${\cal C}$ is a contour in the complex plane that joins $0$ and $\infty$. 
The singularities of the Borel transform, whose knowledge is required to calculate  the integral in (\ref{Bsum}),  are typically determined by obtaining numerically a large number of terms in \Eq{eq:Boreltransf}, 
and  using  a Pad\'e approximant extrapolation \cite{Heller:2015dha,Heller:2016rtz}. Using such a  procedure $g(w)$ was obtained  in the form of a  trans-series \cite{Heller:2015dha}
\be
\label{eq:trans}
 g(w) =\sum_{m=0} \sigma^m g^{(m)}(w)\,, 
\ee
where $\sigma$ is a complex expansion parameter  and 
\be\label{gmn}
g^{(m)}(w)= [\zeta(w)]^m \sum_n f_n^{(m)} w^{-n}
\ee
is itself an asymptotic series. Note that   $g^{(0)}(w)=g_{\rm hydro}(w)$. The quantity $\zeta(w)$, which captures exponential corrections,  can be determined by perturbing the solution around the hydrodynamic gradient expansion \cite{Basar:2015ava}. When applied to the solution of  \Eq{eq:dif_gw}, this yields \cite{Blaizot:2019scw}
\be
\zeta(w) = \rme^{-w}w^{b-2 a+1}= \rme^{-w}w^{a_0-a_1}.
\ee
The same quantity $\zeta(w)$ plays a role in the Borel summation. Indeed the Borel transform in \Eq{eq:Boreltransf} exhibits a branch cut on the positive real axis, which  forces the integration contour ${\cal C}$ in Eq.~(\ref{Bsum}) to be deformed into the complex plane. This  generates an imaginary part in the Borel sum,  which can be shown to be proportional to $\zeta(w)$ \cite{Basar:2015ava}. For  the Borel sum to be real, this imaginary part has to cancel against the next order in the trans-series in \Eq{eq:trans}, which involves the imaginary part of the complex expansion parameter $\sigma$. Such cancellations between coefficients of different members of the trans-series is an illustration of  the well-known resurgence phenomenon (see. e.g. \Refs{Aniceto:2018bis,Basar:2015ava}), to which we shall return. 

\begin{figure}
\begin{center}
\includegraphics[width=0.45\textwidth] {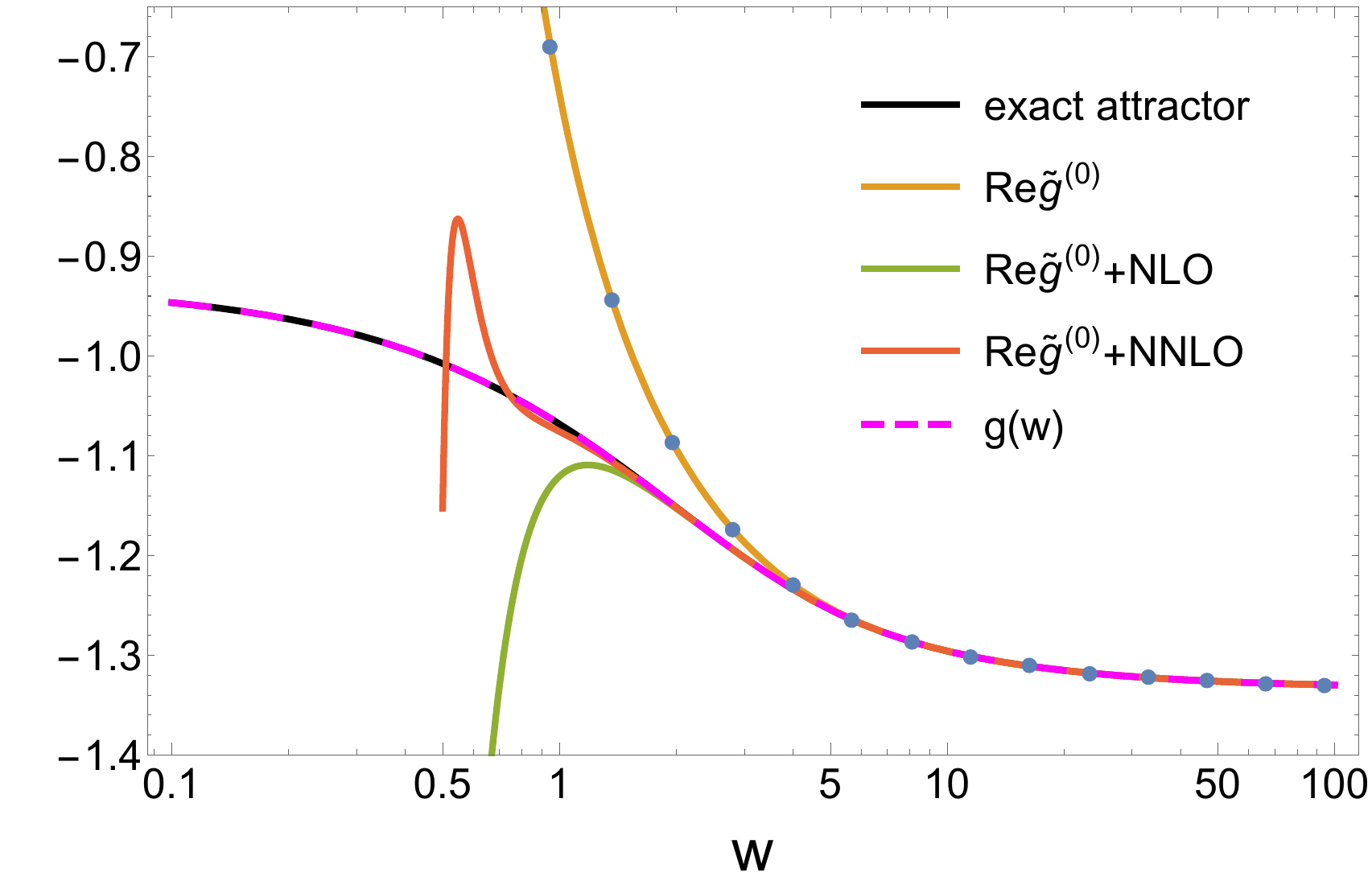}
\caption{ (Color online) 
Borel sums of the trans-series at zero, one, two and infinite order in the expansion (\ref{eq:realg}), in comparison with the exact attractor. The \emph{analytical} Borel sum of the hydrodynamic gradient expansion obtained from Eq.~(\ref{eq:ghydro2b}) (the orange curve) is compared with the Borel sum obtained from the standard method of numerically summing a large number (700) of terms in the series in Eq.~(\ref{eq:ghydro2}) and extrapolating with the help of Pad\'e approximants (blue dots).
\label{fig:resurgence}
}
\end{center}
\end{figure}

Since we have the analytical solution (\ref{eq:gsol2}) at hand, together with the  known asymptotic properties of the confluent hypergeometric functions \cite{10.5555/1098650}, one can verify very explicitly the features of the solution that we have just recalled, without going through a numerical Borel summation (while at the same time allowing for a precise test of this procedure -- see Fig~\ref{fig:resurgence}). The asymptotic forms of the confluent hypergeometric functions $M(a,b,w)$ and $U(a,b,w)$ contain both an exponential factor $e^w$ and a non-analytic power factor  $w^\alpha$, with  $\alpha$ determined by the parameters $a$ and $b$  \cite{10.5555/1098650}. In the analytical solution \Eq{eq:gsol2} these factors naturally combine into $\zeta(w)$, yielding the asymptotic expression 
\begin{align}
\label{eq:Masymp}
&g(w)=g_+ - w \nonumber\\
&+   \frac{   w {\cal F}(-a,b-a,w) - a\sigma \frac{\zeta(w)}{w}  {\cal F}(1+a,1+a-b,-w)   }
{ {\cal F}(1-a,b-a,w) + \sigma \frac{\zeta(w)}{w} {\cal F}(a,1+a-b,-w)  },
\end{align}
where
\be\label{sigma}
\sigma 
= \frac{A \Gamma(a)}{\Gamma(b)} + e^{ i\pi a}\frac{ \Gamma(a)}{\Gamma(b-a)}\,,
\ee
and ${\cal F}(a,b,z)$ is  the asymptotic series
\be\label{calF}
{\cal F}(a,b,z) \equiv \sum_{n=0} \frac{\Gamma(a+n) \Gamma(b+n)}{\Gamma(a) \Gamma(b)} \frac{z^{-n}}{n!}\,.
\ee
This series is Borel summable. 
The Borel transform  is a Gauss hypergeometric function, $\mathcal{B}[{\cal F}(a,b,z)]={}_2F_1(a,b,1,z)$, which has a branch cut running from $z=1$ to $\infty$. The Borel sum  $\tilde {\cal F}$  is  the inverse Laplace transform of  ${}_2F_1(a,b,1,z)$. Its analytical expression reads~\cite{PhysRevA.32.1341}
\begin{align}\label{tildeF}
\frac{\t{\cal F}(a,b,z)}{\pi \csc[(a-b)\pi]} =& -\frac{e^{-i a \pi} z^a M(a,1+a-b,-z)}{\Gamma(b)\Gamma(1+a-b)}\nonumber\\
& + \frac{e^{-i b \pi} z^b M(b,1-a+b,-z)}{\Gamma(a)\Gamma(1-a+b)}\,.
\end{align}

Clearly, the expansion of Eq.~(\ref{eq:Masymp}) in powers of $\sigma$ generates a trans-series of the form of \Eq{eq:trans}. The leading order term ($O(\sigma^0)$) is the hydrodynamic gradient expansion,
\be 
\label{eq:ghydro2}
 g^{(0)}(w) = g_+ -  w\left(1-\frac{{\cal F}(-a,b-a,w)}{{\cal F}(1-a,b-a,w)}\right)\,,
\ee
whose Taylor expansion coefficients can be verifed to coincide with those obtained by the method indicated after  \Eq{eq:hydroexp}. Moreover, the Borel summability of ${\cal F}$, and the availability of the analytical expression of the Borel sum $\tilde {\cal F}$,  suggests the following analytical expression for the Borel sum of the hydrodynamic gradient expansion \be 
\label{eq:ghydro2b}
\tilde g^{(0)}(w) = g_+ -  w\left(1-\frac{\tilde {\cal F}(-a,b-a,w)}{\tilde{\cal F}(1-a,b-a,w)}\right)\,,
\ee
obtained by substituting ${\cal F}$ by $\tilde {\cal F}$ in Eq.~(\ref{eq:ghydro2}), with $\tilde {\cal F}$ given in Eq.~(\ref{tildeF}).   In fact this is an exact procedure. Indeed one can verify that the substitution of ${\cal F}$ by $\tilde{\cal F}$ in Eq.~(\ref{eq:Masymp}) reconstructs the exact solution. 
The higher order asymptotic series $g^{(m)}$ in the trans-series (\ref{eq:trans})  can be determined from Eq.~(\ref{eq:Masymp}) in terms of the function ${\cal F}(a,b,z)$ as well, and their Borel sums $\tilde g^{(m)}$  obtained similarly by substituting  ${\cal F}$  by $\tilde {\cal F}$ in the corresponding expressions. For instance, the analytical Borel sum of $ g^{(1)}(w)$ reads
\begin{align}
&\t g^{(1)}(w) = -\zeta(w)\frac{\t  {\cal F}(-a,b-a,w)}{\t {\cal F}(1-a,b-a,w)} \cr
\quad&\times\left(\frac{a\t {\cal F}(1+a,1+a-b,-w)}{w \t {\cal F}(-a,b-a,w)}+\frac{\t {\cal F}(a,1+a-b,-w)}{\t {\cal F}(1-a,b-a,w)}\right).\cr
\end{align}

 The Borel sum of the hydrodynamic expansion has an imaginary part, related, as mentioned above, to the branch cut of the Borel transform. This imaginary part is cancelled by the real part of the first correction, proportional to the imaginary part of $\sigma$. More precisely, we have 
 \be
 {\rm Im}\t g^{(0)}=S_1 {\rm Re}\t g^{(1)}/2, \qquad S_1=-2{\rm Im}\sigma,
 \ee
 where $S_1$, a so-called Stokes constant, is proportional to the imaginary part of $\sigma$ (and hence independent of the initial condition). 
This cancellation of imaginary parts is one manifestation of the resurgence phenomenon~\cite{Aniceto:2013fka}.  
Resurgence also implies relations between  the  coefficients of the trans-series, which can be easily verified. For instance, the $n^{\rm th}$ order term in the hydrodynamic gradient expansion is related to higher order terms in the trans-series by
\be
2\pi f_n^{(0)}\sim S_1 \Gamma(n+b-2a+1) [f_0^{(1)}+f_1^{(1)}/(n+b-2a)+\ldots]
\ee
 (see e.g. Eq. (39) in \Ref{Basar:2015ava}). Already the first two terms in this expansion yield an excellent approximation to $f_n^{(0)}$ for not too small $n$.
But perhaps the most powerful test  of  the resurgence properties is to consider the complete trans-series solution~\cite{Marino:2008ya,Aniceto:2013fka}, 
\be
\label{eq:realg}
g(w) ={\rm Re} \tilde g^{(0)}(w) + \sigma_{\cal R} {\rm Re}\tilde g^{(1)}(w) 
+ (\sigma_{\cal R}^2-\frac{S_1^2}{4}) {\rm Re}\tilde g^{(2)}(w)
+ \cdots
\ee
where 
$\sigma_{\cal R}={\rm Re}\,\sigma$ is the real part of the $\sigma$. This equation is non-trivial. It encompasses all cancellations of imaginary parts, according to the resurgence relations.  \Fig{fig:resurgence} displays  the trans-series solution  for the attractor. The convergence towards the exact solution is illustrated by adding to the Borel sum of the hydrodynamic gradient expansion $\tilde g^{(0)}$, the contributions proportional to $\tilde g^{(1)}$  (NLO), and to $\tilde g^{(2)}$ (NNLO). Including all orders in the trans-series amounts to substitute ${\cal F}$ by $\tilde {\cal F}$ in Eq~(\ref{eq:Masymp}), which, as we have mentioned, reproduces the exact solution (see Fig.~\ref{fig:resurgence}). 

Further examples of  mathematical relations that can be explored with the analytic solution  will be discussed in \cite{Blaizot_Yan}. Let us just mention here the rearrangement of the trans-series,  leading to the so-called ``transasymptotic matching''  (see e.g. \cite{Basar:2015ava,behtash2020transasymptotics} and references therein). In essence, the procedure consists in summing the exponential corrections in front of each power of $w^{-n}$, thereby  increasing the accuracy of the trans-series in the region of not so large $w$ where these  corrections become of the same order of magnitude as the corresponding terms in the gradient expansion. One then finds that the resulting trans-series  
\be
g(w)=\sum_{n=0}^\infty \frac{1}{w^n}  \sum_{m=0}^\infty g^{(m)}_n \sigma^m \zeta^m.
\ee
has here a simple structure. Indeed, the coefficient of each order $n$ is a polynomial of order $n+1$ in $\zeta$, as can be easily deduced from Eq.~(\ref{eq:Masymp}).  For instance at order $1/w$ we get
\be\label{effectiveeta}
\frac{1}{w}  \left(  c_0 b_1-2 c_0 b_1 \sigma \zeta  + \sigma^2 \zeta^2   \right).
\ee
We recognize in the first term the leading order in the gradient expansion, with the coefficient $b_1$ proportional to the viscosity (see the discussion following Eqs.~(\ref{eq:Lequ})). Following \cite{Behtash:2018moe}, one may then interpret the whole collection of  terms within  the brackets in Eq.~(\ref{effectiveeta}) as an effective viscosity. This effective viscosity, normalized to the standard value, decreases as $w$ decreases. This is the same behavior as that observed in \cite{Blaizot:2017ucy}, but for a completely different reason: there the renormalization of the viscosity originates from the contribution of the higher $\L_n$ moments that are neglected in the two-moment truncation. This observation points to the fact that various factors, of different dynamical origins, may contribute to modify the value of the effective viscosity extracted in heavy ion collisions. 

As a final remark, we note that the attractor solution (\ref{eq:expgatt}) receives exponential corrections from all the ``non-hydrodynamic'' sectors of the trans-series (\ref{eq:trans}), an observation also made in \cite{Heller:2015dha}. From that perspective, we find the often used terminology ``hydrodynamic attractor"  somewhat misleading. As we have emphasized in this letter, we prefer to view the attractor, in the present context, as the particular solution that connects two fixed points of the non linear equation (\ref{eq:dif_gw}) for the pressure asymmetry. The two fixed points  are associated with different physics: one corresponds to hydrodynamics, the other to the collisionless regime. The requirement of connecting these two different regimes is reflected in the  intricate mathematical structure of the solution. As we have emphasized, the  particular fixed point structure which emerges naturally from kinetic theory turns out to be shared by all versions of second order hydrodynamics applied to Bjorken flows.  

\emph{Acknowledgements.} --- 
L.Y. is supported in part by National Natural Science Foundation of China (NSFC) 
under Grant No. 11975079.

\bibliography{refsbib}
\end{document}